# Tamm Plasmon Resonance Responsiveness to SARS-CoV-2 Virus-Like Particles

Andrea Rossini[1,2], Fabio Marangi[2], Pietro Bertolotti[2,3], Francesco Scotognella[4], Guglielmo Lanzani[1,2] and Giuseppe Maria Paternò[1,2]*

1. Department of Physics, Politecnico di Milano, Milano, Italy
2. Center for Nano Science and Technology, Istituto Italiano di Tecnologia, Milano, Italy
3. Department of Bioengineering, Politecnico di Milano, Milano, Italy
4. Department of Applied Science and Technology, Politecnico di Torino, Torino, Italy

## Abstract

Bioresponsive optical materials that transduce nanoscale bio–interface events into measurable spectral signals are of growing interest for sensing and antiviral technologies. Here, we show that a Tamm plasmon (TP) device, consisting of a $SiO_2/TiO_2$ distributed Bragg reflector capped with a nanostructured silver layer, exhibits a selective and structure-dependent response to SARS-CoV-2 virus-like particles (VLPs). Upon VLP exposure, the conventional wavelength shift ($\Delta\lambda$) of the TP resonance is minimal, whereas the resonance depth undergoes a systematic attenuation. To capture both spectral and amplitude variations, we introduce a "displacement angle" $\alpha$, defined from the translation vector of the Tamm dip before and after exposure. The angle $\alpha$ increases monotonically with VLP concentration and enables a limit of detection of 1.3 ng/mL. Control experiments with polystyrene nanoparticles of similar size and with heat-denatured VLPs yield negligible changes, indicating sensitivity to the native conformational state of viral surface proteins rather than to generic nanoparticle morphology or bulk refractive index effects. These results establish Tamm plasmon structures as promising bioresponsive platforms for label-free detection and for probing structure-dependent virus–material interactions.

*Correspondence should be addressed to: giuseppemaria.paterno@polimi.it

# 1 Introduction

Designing optical materials that respond selectively to biological nanoscale structures is an emerging frontier in photonic sensing and functional interfaces. Among such platforms, Tamm plasmon (TP) systems, hybrid photonic–plasmonic modes formed at the interface between a metal film and a dielectric Bragg reflector, have attracted growing interest due to their sharp resonances, accessible far-field excitation, and strong interfacial field confinement [1, 2]. These characteristics make TP structures compelling candidates for developing bioresponsive optical materials, where subtle changes at the metal surface can translate into measurable spectral signatures. [3, 4].

While TP devices have been explored in applications such as lasing [5], photodetection [6], and refractive-index sensing [1, 7], their ability to selectively respond to complex biological nanostructures remains underexplored. A central challenge is that the TP mode resides at a buried interface; achieving surface sensitivity typically requires careful engineering of the metallic layer, such as nanoscale roughness or corrugation, to extend the electromagnetic field toward the exposed surface [8, 9]. Recent work demonstrated that such engineered TP devices can differentiate between physiological states of bacteria through combined spectral and amplitude changes in the resonance, revealing their sensitivity to chemically induced perturbations at the metal interface [10]. In that context, the observed TP response resulted from two competing effects: (i) infiltration of the surrounding medium into the DBR/near-surface region, which increases the local effective refractive index and induces a red shift; and (ii) chemical interaction between silver and nucleophilic ligands in the bacterial envelope, leading to oxidative dissolution of the Ag layer and a blue shift associated with changes in carrier density [11-13].

In this work, we investigate whether this TP architecture can function as a selective and structure-dependent responsive material when exposed to virus-like particles (VLPs). VLPs serve as well-defined biological systems that mimic the surface architecture of viruses while remaining non-infectious, making them ideal model analytes for probing material–virus interactions [14]. As a representative and technologically/societal relevant case [3, 15, 16], we use SARS-CoV-2 VLPs to evaluate how intact nanoscale assemblies, distinct from denatured particles or inert colloids, perturb the TP resonance. By analyzing both the spectral shift and the change in resonance depth, we demonstrate that TP devices exhibit a characteristic and concentration-dependent optical response specifically to native VLPs, highlighting their potential as a versatile photonic platform for studying and exploiting bio–nanointeractions at material interfaces.

## 2 Materials and methods

**Fabrication of the device.** Following the procedure reported by S. Normani et al. [10], two aqueous dispersions of $TiO_2$ (GetNanoMaterials, average diameter ≈ 5 nm, 9 % w/V), and $SiO_2$ nanoparticles (LUDOX, average diameter ≈ 5 nm, 7 % w/V) have been prepared. Both dispersions have been sonicated using a tip sonicator (Branson 450 Digital Sonifier) and filtered with a 20 μm PTFE membrane to prevent particle clustering. Prior to deposition, glass substrates (2.5 x 2.5 $cm^2$) have been cleaned by sequential sonication in ultrapure water, acetone and isopropyl alcohol (IPA). Then, the cleaned substrates have been treated with Oxygen plasma for 10 min. (Femto, Diener) to enhance surface cleanliness and adhesion. The Distributed Bragg Reflector (DBR) consists of a fixed number repetition of bilayers, and each layer has been annealed at 420 °C for 20 minutes. Upon completion of the DBRs, 33 nm of silver (Ag) have been deposited on the surface by vacuum thermal evaporation, as in previous works [11, 17]. In the following text, the complete samples, made of the DBR coupled with the silver layer, will be referred to as the Tamm device (TD).

**Analytes preparation and experimental procedure.** SARS-CoV-2 VLPs (Abnova, Catalog No. P667799) consist of four structural proteins: Membrane protein (M), Nucleocapsid protein (N), Spike protein (S), and Envelope protein (E), all derived from HEK293 cells (source: Abnova datasheet). Additional materials used in the sensing experiments included polystyrene nanoparticles (PS NPs) (Merck, Catalog No. 43302), phosphate-buffered saline (PBS) (Fisher Scientific, Catalog No. BP2944100), and agarose (Merck, Catalog No. A9539). Before the experiments, agarose plates have been prepared to serve as substrates for the sensing tests. All aqueous dispersions have been prepared using PBS (0.01 M).

Working dispersions of VLPs have been obtained by serial dilution of a stock dispersion (100 ng/mL) to final concentrations of 0.5, 1, 5, 10 and 50 ng/mL, which will be referred to as VLP_0.5, VLP_1, VLP_5, VLP_10, VLP_50. PBS-only dispersion will be called "VLP_0" in the following part. PS NPs, with a spheric shape and 100 nm of diameter (similar to SARS-CoV-2 [18]), have been diluted in PBS to a working concentration of 10 ng/mL for comparative experiments. For each sensing test, 100 μL of the analyte dispersion (VLP, PS NPs or VLP_0) have been pipetted onto a marked position on an agarose plate. Overnight exposure of the TD was carried out by placing the Ag exposing side on the liquid spot.

To test the sensitivity of the device to the external proteins of the VLP, a 10 ng/mL dispersion has been heated for 40 minutes at 40 °C with a hot plate. The temperature and duration have been selected based on precedent literature [19, 20], aiming to disrupt the conformation of the spike and envelope proteins without compromising the integrity of the viral particle. In subsequent sections, this dispersion is referred to as "VLP_Denatured".

**Optical spectroscopy and displacement angle.** Visible and near-infrared (Vis−NIR) transmission and reflection measurements have been performed at multiple steps during the fabrication of the TD using a UV/VIS/NIR spectrometer (Lambda 1050, Perkin Elmer) equipped with an integrating sphere. Devices have been characterized after the DBR fabrication, after TD fabrication, and following exposure to analyte dispersions, across the 350-850 nm wavelength range.

The figure of merit chosen for the detector is the angle defined by the translation vector connecting the Tamm featured dip before and after the analysis. After extracting the horizontal shift ($\Delta\lambda$) and the vertical attenuation ($\Delta R$ or $\Delta T$) components of this vector, the angle, referred to as "displacement angle" or "α", has been calculated in degrees using the following equation:

$$\alpha(°) = \frac{180}{\pi} * \tan^{-1}\left(\frac{\Delta R}{\Delta \lambda}\right) \quad (1)$$

Once α values have been obtained for different concentrations, the limit of detection (LOD) has been calculated using the following formula [21]:

$$LOD = \frac{3.3 * \sigma_{blank}}{S_{VLP}} \quad (2)$$

in which $\sigma_{blank}$ is the standard deviation for the 0 ng/mL VLP dispersion and S is the slope of the fitting curve in its linear region.

**Scanning Electron Microscopy (SEM).** To optimize DBR fabrication, a scanning electron microscope (Tescan MIRA3) has been used in secondary electron mode to investigate the relationship between the layer thickness and spin-coating parameters via cross-sectional analysis.

**Atomic Force Microscope (AFM).** AFM topography measurements have been carried out to evaluate surface morphology and roughness of the thermally evaporated Ag layer. The images were acquired in Peak Force Tapping Mode by using a ScanAsyst – Air – HPI tip (2 nm nominal tip radius) on a Bruker Dimension Icon XR. The images have been analyzed with Gwyddion software to correct tilt and extract roughness (Rms) on a 10x10 μm area. AFM measurements were also used to evaluate the voids fraction ($V_{void}$) in the Ag layer, by masking the image with Gwyddion software and then using the extracted values of projected area ($A_{projected}$) and of the total area ($A_{total}$) in the following equation:

$$\boldsymbol{Void}\ (\%) = \frac{A_{total} - A_{projected}}{A_{total}} \quad (3)$$

# 3 Results and Discussion

## 3.1 Tamm Device and detection mechanisms

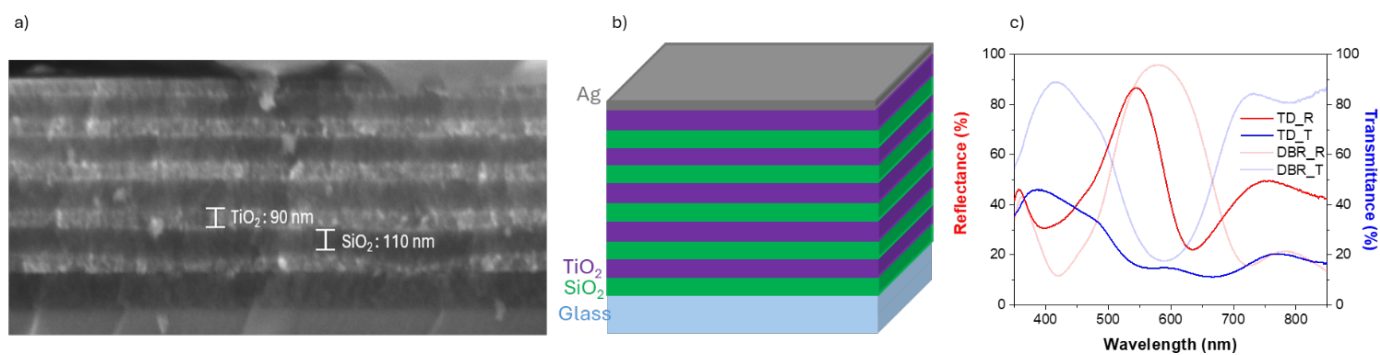

**Figure 1** a) SEM cross-section of a five bilayer DBR(SiO$_2$-110 nm; TiO$_2$-90 nm) deposited onto a glass substrate. b) Scheme of the TD device. c) Transmittance (blue) and reflectance (red) spectra of the DBR (semi-transparent) and of the TD (solid), showing a dip in reflectance at 620 nm.

The DBR cross-section, composed by 10 alternated layers of SiO$_2$ and TiO$_2$, has been characterized using SEM (Fig.1a) to determine layers thicknesses and their homogeneity. The measured thicknesses of the SiO$_2$ and TiO$_2$ layers are respectively 110 nm and 90 nm

Figure 1b,c shows a schematic of the complete TD, and the optical spectra of the DBR (FWHM ≈180 nm) and of the TD. The TD reflectance and transmittance spectra display the characteristic Tamm plasmon feature: a reflectance dip accompanied by a transmittance peak near 620 nm, consistent with previous reports on similar DBR/metal structures [2, 22, 23]. The thickness of the silver layer has been optimized to enhance the Tamm Plasmon response resulting in a more pronounced dip in reflectance. To ensure the sensitivity of the electromagnetic field to the air interface, which is crucial for biosensing applications, the fabrication process has been tailored to introduce nanometric corrugation (Rms= 8.7 nm) in the metallic film, which has been evaluated through AFM measurements The measurements confirmed the formation of silver island-like nanostructures (Fig S1). In fact, the surface morphology of the TD resembles a film of aggregated silver NPs (d$_{mean}$ ~ 50±18 nm). The film is homogeneous despite periodically exposing the DBR structure due to voids (14.4%) deriving from the deposition process.

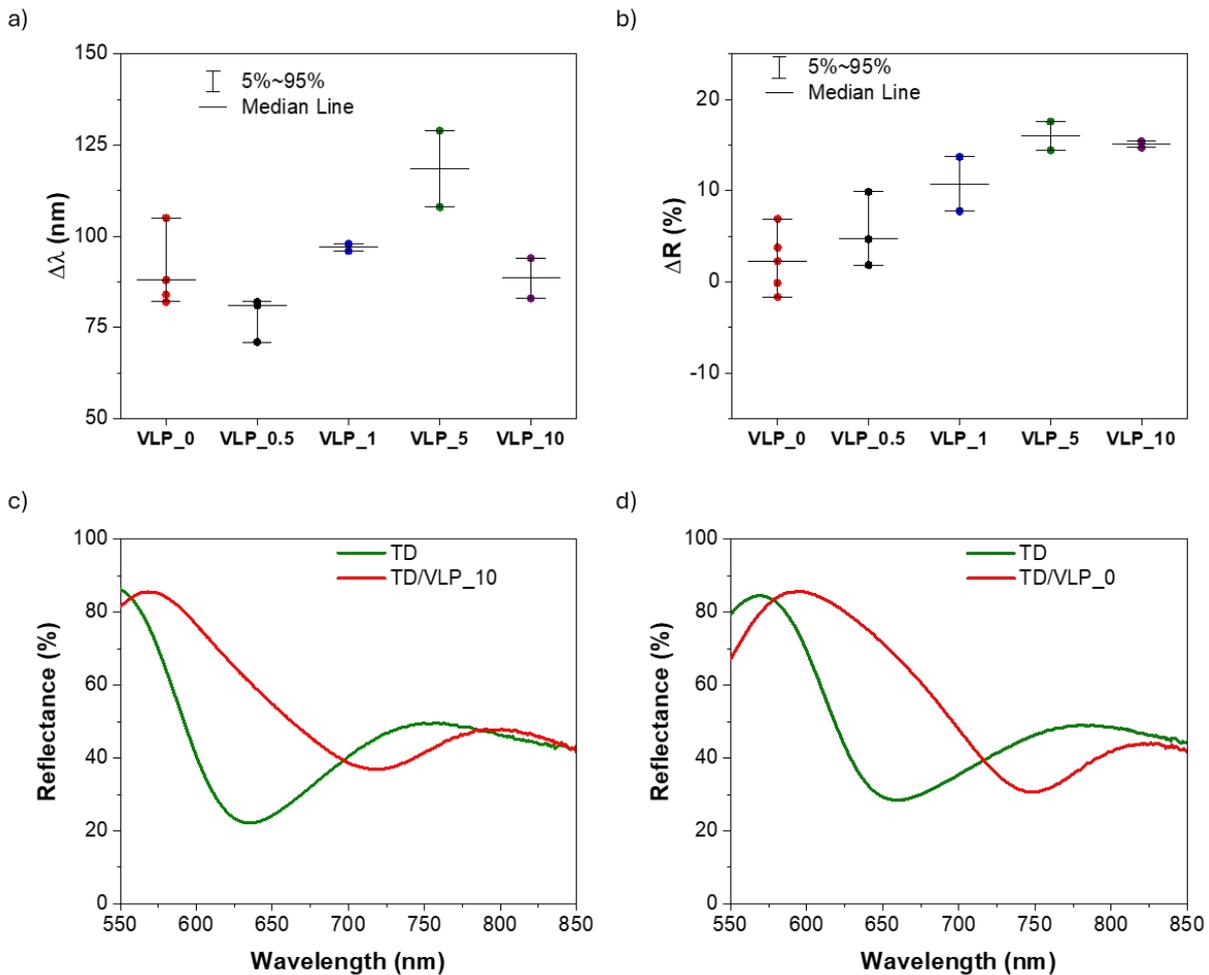

*Figure 2* In panels (a) and (b), the two scatter plots show the wavelength shift (Δλ) and reflectance intensity change (ΔR) of the Tamm feature, respectively, as a function of VLP concentration, in the x-axis. Panels (c) and (d) display the reflectance spectra before and after exposure to VLP_10 and VLP_0. In both cases, the pre-exposure curve is shown in green, while the post-exposure curve is in red.

Following fabrication and initial optical characterization, the TDs were exposed to SARS-CoV-2 VLPs at different concentrations. As a first step, we compared the wavelength shift (Δλ) of the Tamm resonance between the sample with and without VLP, exploiting the same figure of merit used previously for bacteria detection, namely the blue shift [10], while considering the biological and chemical differences between bacteria and virus. Several studies have reported that viral surface proteins can adsorb onto silver and, in some cases, induce oxidative dissolution [24-27]. However, the amount of $Ag^+$ ions released is generally considered too low to generate a pronounced antiviral effect. This picture is consistent with our experimental data in Fig. 2a, where Δλ does not show a trend comparing the cases with and without SARS-CoV-2 and does not provide a reliable metric for detecting the presence of SARS-CoV-2 VLPs with the TD. The absence of a Δλ reduction, a blue shift, in the experiment containing the VLP suggests a weaker oxidative dissolution response compared to the bacterial case [10]. Given the limited sensitivity of Δλ, we focused instead on the vertical component of the Tamm feature translation, ΔR% or ΔT%, which reflects changes in resonance intensity. All data discussed here are extracted from reflectance spectra, where the Tamm dip is more obvious to infer and thus constitutes a more sensitive observable. As shown in Fig. 2b, increasing VLP concentration leads to a systematic rise in the reflectance minimum, corresponding to a shallower dip. To interpret this behavior, we consider two aspects: (i) the virus–silver interaction and (ii) the attenuation of the Tamm resonance. For the first point, we refer to the work of Paternò et al. [13], which investigates the photophysical and structural modifications of silver nanoparticles upon interaction with bacteria. Their analysis revealed damping and broadening of the localized surface plasmon resonance (LSPR) peak, with a loss of coherent oscillation attributed to changes in free-electron density and Ag amorphization. Several similarities can be drawn between their system and ours: (i) adsorption of biological species at the metal interface, (ii) oxidative dissolution of silver triggered by the interaction, and (iii) a comparable exposed silver morphology (nanoplates and an island-like microstructure, see Fig. S1). These elements support a mechanistic comparison between the two cases. Moreover, the behavior observed for the LSPR peak in Ref. [13] can be expected to affect the TP feature as well. Indeed, Auguié et al. [2] have shown that, for Tamm resonances, increased optical losses or degraded interface quality reduce the resonance depth and broaden the linewidth, which are optical signatures of higher modal damping. Considering our data and these previous studies, the attenuation and broadening of the Tamm dip observed in Fig. 2c can be attributed to increased damping and deviation from critical coupling conditions, likely driven by changes in the free-electron population, and hypothetically a structural modification of the silver microstructure, following virus interaction, similar to the one observed in A. Ross work [13]. In Fig. 2c,d, the TD reflectance spectra before and after exposure to VLP_0 ng/mL and VLP_10 ng/mL can be compared to support this interpretation. The spectral shift (88 nm for 0 ng/mL and 83 nm for 10 ng/mL VLP) is similar in both cases, while the difference in ΔR% is evident.

## 3.2 VLP concentrations test

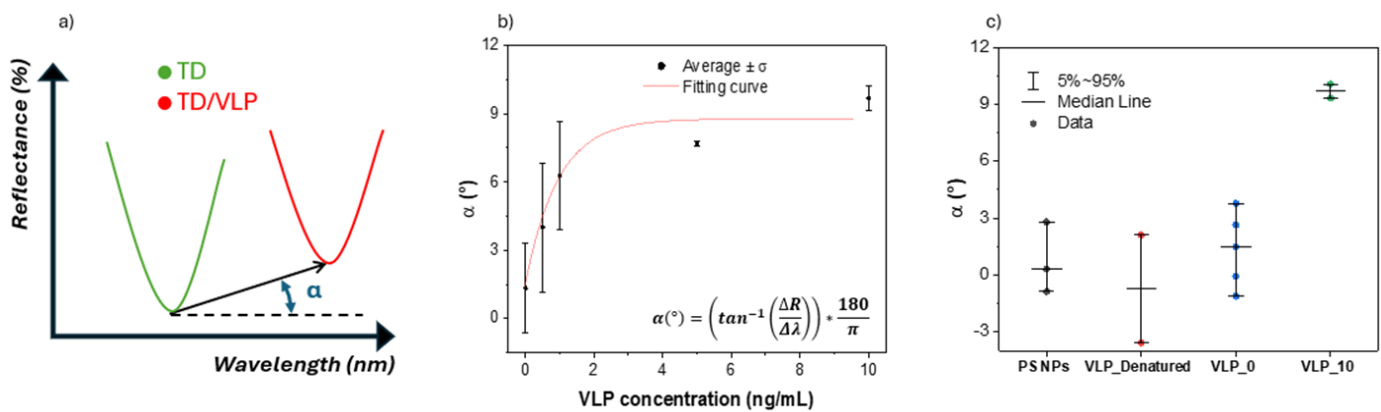

*Figure 3* (a) Schematic representation of the parameters used to calculate the displacement angle α, derived from the translation vector connecting the Tamm dip before and after exposure. This angle serves as the sensing output. (b) Graph of α vs VLP concentration, showing the average value and standard deviation of the tested concentrations: VLP_0 (5 tests), VLP_0.5 (3 tests), VLP_1 (2 tests), VLP_5 (2 tests), and VLP_10 (2 tests). The red curve represents the fitting function, chosen as an exponential decay. The equation used to calculate α is reported in the lower right corner of the plot. (c) scatter plot of the displacement angle (α) with respect to the VLP concentration, comparing experimental data obtained from different dispersions: 10 ng/mL of PS NPs, VLP_Denatured", VLP_0 and VLP_10.

To capture both spectral position and amplitude changes in a single, robust metric, the angle α of the translation vector, connecting the Tamm dip before and after exposure, has been extrapolated as reported in the scheme in Fig. 3a. This angle incorporates both the horizontal shift (Δλ) and vertical attenuation (ΔR) components, making it sensitive to dielectric loading, increased optical loss, surface scattering, and changes at the metal interface. In Fig 3b, α provides a more comprehensive and reliable indicator of TD response to SARS-CoV-2 VLPs than Δλ and ΔR% alone. This graph of displacement angle versus VLP concentrations has been constructed by calculating the average α value and standard deviation for each tested concentration. Given the nature of the interaction mechanism, the curve is expected to exhibit an initial logarithmic slope followed by saturation at a limiting value. Accordingly, the experimental data, α vs VLP concentration, have been fitted using the following equation:

$$y = y_0 - A * e^{-\frac{x}{t}} \tag{4}$$

where y is the displacement angle, x is the VLP concentration, and $y_0$, A, t are the fitting parameters. The limit of detection (LOD) of the device has been determined to be 1.3 ng/mL, indicating that the sensor can reliably detect the presence of SARS-CoV-2 VLP concentrations above this threshold. To estimate a saturation limit, dispersions at 50 ng/mL have been tested; however, the displacement angle could not be extracted because the Tamm feature had become indistinct in the output spectrum (Fig. S2). Thus, the saturation value should not be far from the tested concentration of 10 ng/mL.

To validate the observed phenomenon, several control tests were performed. Following procedures from a previous study [10], a bare DBR and a silver layer were tested with PBS and VLP dispersions. No significant differences were observed between spectra recorded with and without VLP exposure (Fig. S3), highlighting the essential role of the silver layer and, hence, of the Tamm plasmon in this sensing mechanism. Another test investigated whether inert nanoparticles with a similar dimension can influence the sensor output only due to presence of the nanoparticles or by changing the total refractive index [28]. A dispersion containing PS NPs with a nominal diameter of 100 nm was tested, and the results, shown in Fig. 3c, revealed no differences compared to the PBS solution. This confirms that the device is not sensitive to particle morphology and change in the refractive index alone, but rather to the specific interaction between the analyte and the silver interface.

A final test was conducted to evaluate whether the biological state of the viral surface proteins influences the sensor response, with potential implications for drug-screening applications. To this end, we denatured the VLP proteins via heat treatment, and the resulting "VLP_Denatured" dispersion was then interfaced with our devices. The results (Fig. 3c) show no statistically significant difference between this sample and the control (0 ng/mL). This indicates virtually no interaction between the silver layer and the capsid proteins when they are denatured and therefore nonfunctional. Since our TD devices are sensitive to the biologically active state of the viral envelope proteins, they could potentially serve as a platform for evaluating the effects of treatments or drugs that target these surface structures.

## 4 Conclusion

In this study, the TDs have been validated as sensors to detect the SARS-CoV-2 VLPs, showing distinct response at different concentrations and a LOD of 1.3 ng/mL. Compared to previous literature, the observed shift and broadening of the TP resonance after exposure (captured in the displacement angle) appear to be associated with increased optical energy loss and modal damping, likely caused by interactions between the silver layer and the VLPs. Control tests performed with similarly sized particles (PS NPs) and denatured VLPs support this interpretation. Moreover, the output from the latter shows a possible application of the device for drug screening applications. While other sensors reported in the literature can achieve lower detection limits or target specific functional groups [29, 30] the simplicity and ease of fabrication of the TD make it an attractive platform from both practical and technological perspectives.

## Supporting Information

Supporting Information is available from the Wiley Online Library or from the author.

## Acknowledgements

This work is co-funded by the European Union (ERC, EOS, 101115925). Views and opinions expressed are, however, those of the author(s) only and do not necessarily reflect those of the European Union or the European Research Council. Neither the European Union nor the granting authority can be held responsible for them.

## Conflict of Interest

The authors declare no conflict of interest

## Data Availability

The data that support the findings of this study are openly available in Zenodo at DOI: 10.5281/zenodo.18064645.

## References

[1]     O. Haidar *et al.*, "Enhancing Tamm Plasmon Sensor Performance Using Nanostructured Gold Grating and Porous Materials," *IEEE Sens J*, vol. 24, no. 13, pp. 204r52–20459, Jul. 2024, doi: 10.1109/JSEN.2024.3399315.

[2]     B. Auguié, A. Bruchhausen, and A. Fainstein, "Critical coupling to Tamm plasmons," *Journal of Optics (United Kingdom)*, vol. 17, no. 3, Feb. 2015, doi: 10.1088/2040-8978/17/3/035003.

[3]     T. Tene, M. Guevara, P. Romero, A. Guapi, L. Gahramanli, and C. Vacacela Gomez, "SARS-CoV-2 detection by surface plasmon resonance biosensors based on graphene-multilayer structures," *Front Phys*, vol. 12, 2024, doi: 10.3389/fphy.2024.1503400.

[4]     S. Wallace *et al.*, "Multiplexed Biosensing of Proteins and Virions with Disposable Plasmonic Assays," *ACS Sens*, vol. 8, no. 9, pp. 3338–3348, Sep. 2023, doi: 10.1021/acssensors.2c02238.

[5]     J. Zhang *et al.*, "Excitation of Tamm plasmon polariton in ultrathin metals," Sci. Adv. 11, 2025, doi: 10.1126/sciadv.adz0106

[6]     W. Liang, Y. Dong, L. Wen, and Y. Long, "Silicon-based planar devices for narrow-band near-infrared photodetection using Tamm plasmons," *Nanophotonics*, vol. 13, no. 16, pp. 2961–2970, Jul. 2024, doi: 10.1515/nanoph-2024-0062.

[7]     S. Sidhireddy and N. Ashok, "Design and Analysis of a Tri-layer Reflective Structure–Based Tamm Plasmon Sensor for Cancer Cell Detection," *Plasmonics*, Oct. 2025, doi: 10.1007/s11468-025-02985-7.

[8]     S. Normani, F. F. Carboni, G. Lanzani, F. Scotognella, and G. M. Paternò, "The impact of Tamm plasmons on photonic crystals technology," *Physica B Condens Matter*, vol. 645, Nov. 2022, doi: 10.1016/j.physb.2022.414253.

[9]     A. Rossini *et al.*, "Plasmonic Tamm Resonance in a Conjugated-Polymer Biointerface for Efficient Cell Photostimulation," *Adv Opt Mater*, 2025.

[10]    S. Normani *et al.*, "Tamm Plasmon Resonance as Optical Fingerprint of Silver/Bacteria Interaction," *ACS Appl Mater Interfaces*, vol. 15, no. 23, pp. 27750–27758, Jun. 2023, doi: 10.1021/acsami.3c05473.

[11]    G. M. Paternò *et al.*, "Integration of bio-responsive silver in 1D photonic crystals: Towards the colorimetric detection of bacteria," *Faraday Discuss*, vol. 223, pp. 125–135, Oct. 2020, doi: 10.1039/d0fd00026d.

[12]  S. Normani, N. Dalla Vedova, G. Lanzani, F. Scotognella, and G. M. Paternò, "Bringing the interaction of silver nanoparticles with bacteria to light," Jun. 01, 2021, *American Institute of Physics*. doi: 10.1063/5.0048725.

[13]  G. M. Paternò *et al.*, "The Impact of Bacteria Exposure on the Plasmonic Response of Silver Nanostructured Surfaces," Jan. 2021, doi: 10.1063/5.0042547.

[14]  S. Nooraei *et al.*, "Virus-like particles: preparation, immunogenicity and their roles as nanovaccines and drug nanocarriers," Dec. 01, 2021, *BioMed Central Ltd*. doi: 10.1186/s12951-021-00806-7.

[15]  Y. Saad, M. H. Gazzah, K. Mougin, M. Selmi, and H. Belmabrouk, "Sensitive Detection of SARS-CoV-2 Using a Novel Plasmonic Fiber Optic Biosensor Design," *Plasmonics*, vol. 17, no. 4, pp. 1489–1500, Aug. 2022, doi: 10.1007/s11468-022-01639-2.

[16]  T. L. Meister *et al.*, "Nanoscale copper and silver thin film systems display differences in antiviral and antibacterial properties," *Sci Rep*, vol. 12, no. 1, Dec. 2022, doi: 10.1038/s41598-022-11212-w.

[17]  G. M. Paternò *et al.*, "Hybrid one-dimensional plasmonic-photonic crystals for optical detection of bacterial contaminants," *Journal of Physical Chemistry Letters*, vol. 10, no. 17, pp. 4980–4986, Sep. 2019, doi: 10.1021/acs.jpclett.9b01612.

[18]  E. Mosidze *et al.*, "Silver Nanoparticle-Mediated Antiviral Efficacy against Enveloped Viruses: A Comprehensive Review," May 01, 2025, *John Wiley and Sons Inc*. doi: 10.1002/gch2.202400380.

[19]  B. Pastorino, F. Touret, M. Gilles, X. de Lamballerie, and R. N. Charrel, "Evaluation of heating and chemical protocols for inactivating SARS-CoV-2," Apr. 11, 2020. doi: 10.1101/2020.04.11.036855.

[20]  A. Gamble, R. J. Fischer, D. H. Morris, C. K. Yinda, V. J. Munster, and J. O. Lloyd-Smith, "Heat-Treated Virus Inactivation Rate Depends Strongly on Treatment Procedure: Illustration with SARS-CoV-2," *Appl Environ Microbiol*, vol. 87, no. 19, pp. 1–9, Sep. 2021, doi: 10.1128/AEM.00314-21.

[21]  "INTERNATIONAL CONFERENCE ON HARMONISATION OF TECHNICAL REQUIREMENTS FOR REGISTRATION OF PHARMACEUTICALS FOR HUMAN USE ICH HARMONISED TRIPARTITE GUIDELINE VALIDATION OF ANALYTICAL PROCEDURES: TEXT AND METHODOLOGY Q2(R1)."

[22]  S. Sudha Maria Lis, S. Pandit, S. Patra, D. Banerjee, and B. N. Shivakiran Bhaktha, "Tamm Mode-Aided Amplified Spontaneous Emission in One-Dimensional Photonic Crystal Super-Tamm Structure," in *CLEO: Fundamental Science, CLEO:FS 2023*, Optical Society of America, 2023. doi: 10.1364/CLEO_FS.2023.FF2D.4.

[23]  L. Ferrier *et al.*, "Tamm plasmon photonic crystals: From bandgap engineering to defect cavity," *APL Photonics*, vol. 4, no. 10, Oct. 2019, doi: 10.1063/1.5104334.

[24]  J. Hilton *et al.*, "The role of ion dissolution in metal and metal oxide surface inactivation of SARS-CoV-2," *Appl Environ Microbiol*, vol. 90, no. 2, Feb. 2024, doi: 10.1128/aem.01553-23.

[25]  M. Sahihi and J. Faraudo, "Computer Simulation of the Interaction between SARS-CoV-2 Spike Protein and the Surface of Coinage Metals," *Langmuir*, vol. 38, no. 48, pp. 14673–14685, Dec. 2022, doi: 10.1021/acs.langmuir.2c02120.

[26]  T. Miclăuş *et al.*, "Dynamic protein coronas revealed as a modulator of silver nanoparticle sulphidation in vitro," *Nat Commun*, vol. 7, Jun. 2016, doi: 10.1038/ncomms11770.

[27]  G. Gupta *et al.*, "Silver nanoparticles with excellent biocompatibility block pseudotyped SARS-CoV-2 in the presence of lung surfactant," *Front Bioeng Biotechnol*, vol. 10, Dec. 2022, doi: 10.3389/fbioe.2022.1083232.

[28] M. C. de Oliveira *et al.*, "Unraveling the Intrinsic Biocidal Activity of the SiO2-Ag Composite against SARS-CoV-2: A Joint Experimental and Theoretical Study," *ACS Appl Mater Interfaces*, vol. 15, no. 5, pp. 6548–6560, Feb. 2023, doi: 10.1021/acsami.2c21011.

[29] Y.-R. Li *et al.*, "Direct detection of virus-like particles using color images of plasmonic nanostructures," *Opt Express*, vol. 30, no. 12, p. 22233, Jun. 2022, doi: 10.1364/oe.461428.

[30] J. Kulanthaivel, V. Hitaishi, and N. Ashok, "A Tamm-Fano resonance glucose sensor based on Cu and distributed bragg reflector plasmonic coupling interface in the near-infrared regime," *Opt Quantum Electron*, vol. 56, no. 9, p. 1538, Sep. 2024, doi: 10.1007/s11082-024-07396-2.

# Supplementary material of "Bioresponsive Tamm Plasmon Devices for Selective Optical Detection of SARS-CoV-2 Virus-Like Particles"

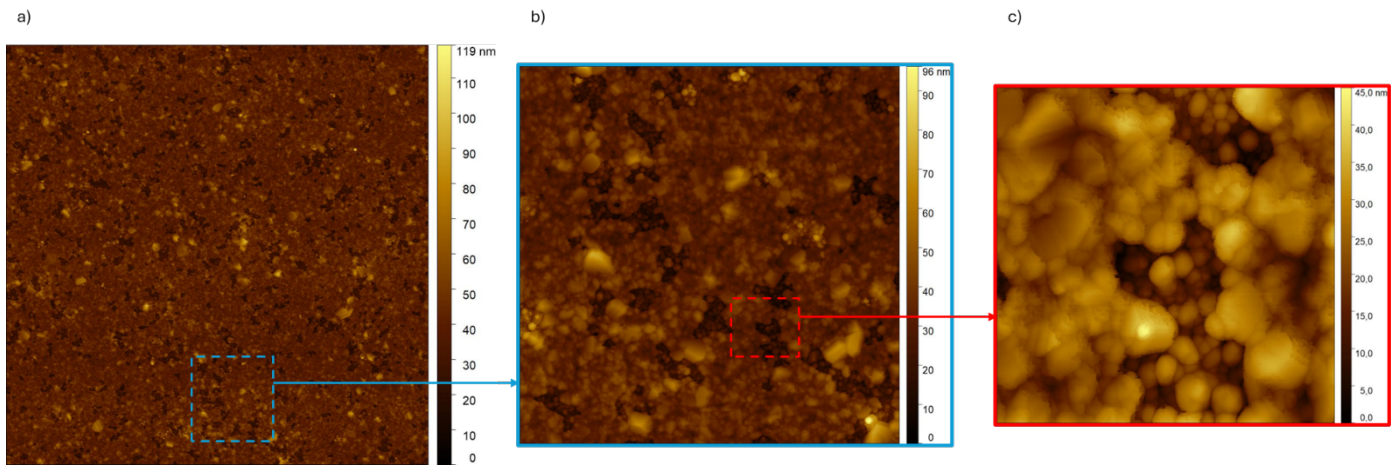

**Figure S1** *Characterization of the silver (Ag) nanostructure through AFM analysis. (a) AFM image of the Tamm device top surface, 10x10 μm area, highlighting the non-bulk nanostructure of the Ag film. (b) magnification 2.5x2.5 μm (c) Magnification 0.5x0.5 μm, showing one of the voids and the island-like structure of the Ag film.*

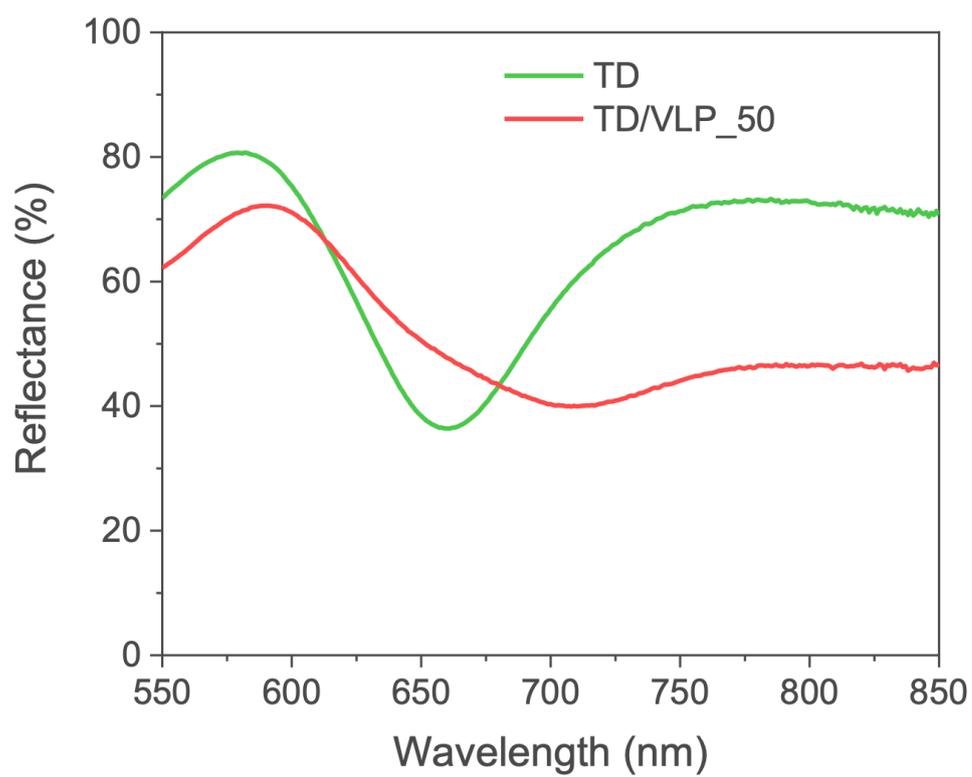

***Figure S2**  The reflectance spectra of the TD before (green) and after (red) exposure to 50 ng/mL of VLPs.*

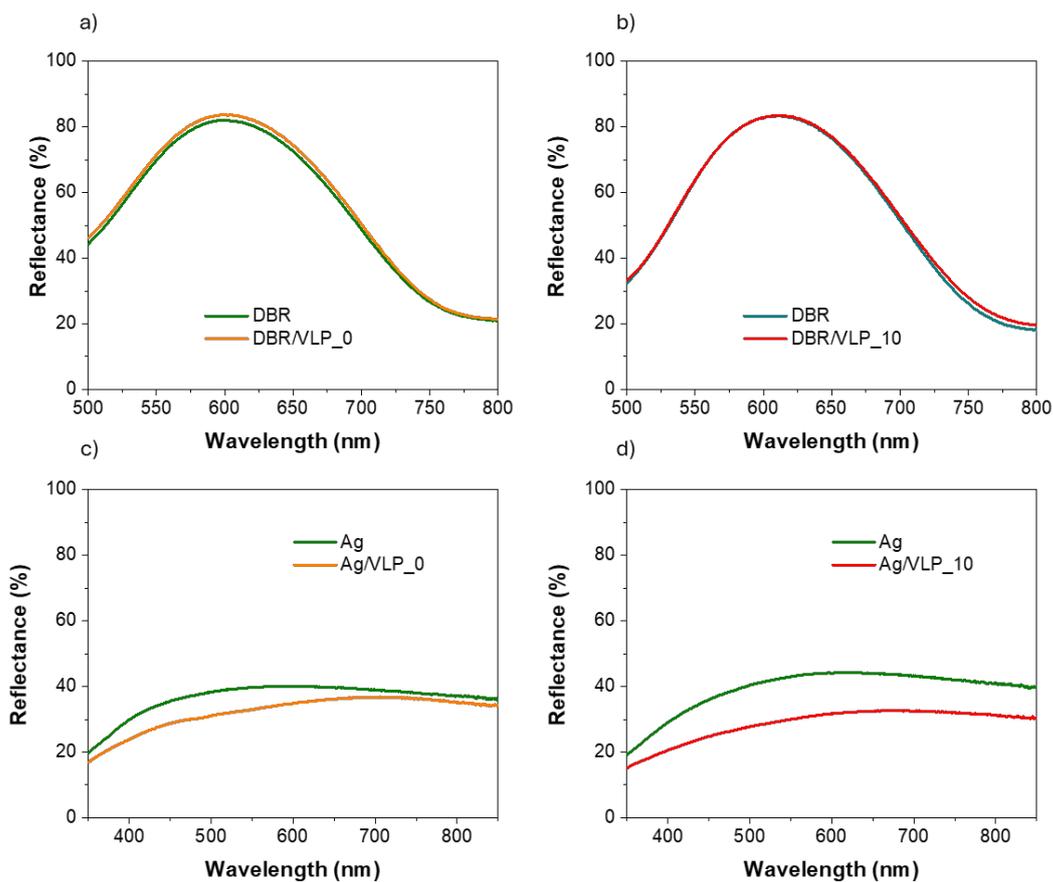

***Figure S3*** *Panel (a) and (b) report the reflectance spectra of the DBR before (green) and after exposure to VLP_0 (0 ng/mL in orange) and VLP_10 (10 ng/mL in red). Panel (c) and (d) display the reflectance spectra of a 33 nm thick layer of Ag, deposited on a glass substrate, before (green) and after exposure to VLP_0 (0 ng/mL in orange) and VLP_10 (10 ng/mL in red).*

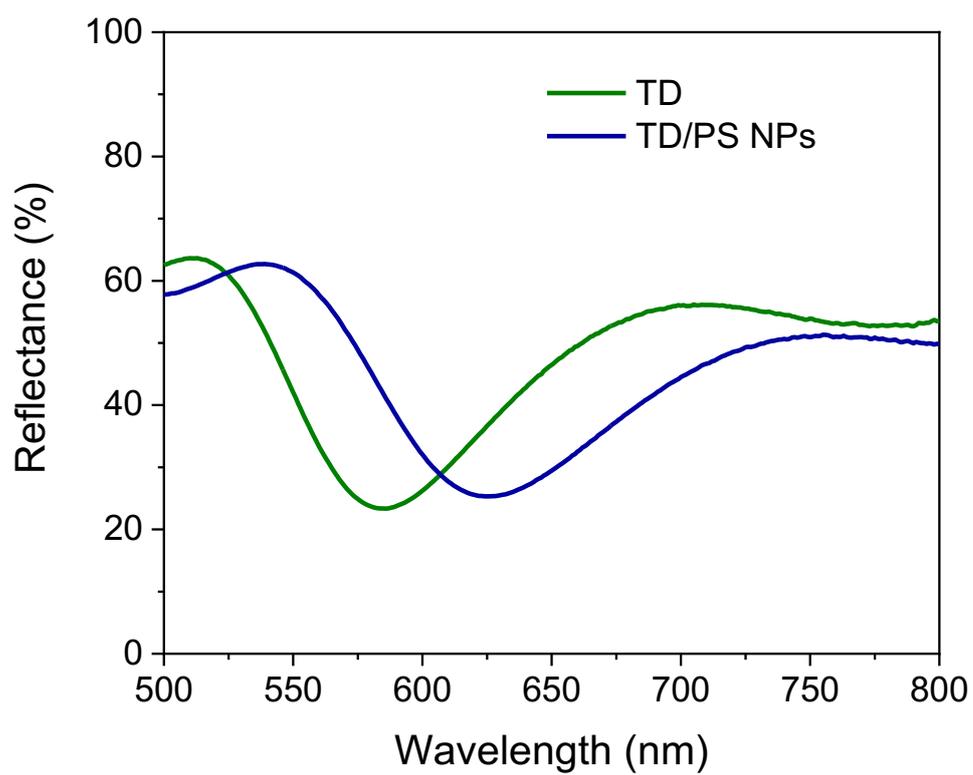

*Figure S4* *The reflectance spectra of the TD before (green) and after (blue) exposure to 10 ng/mL of Polystyrene nanoparticles (PS NPs).*

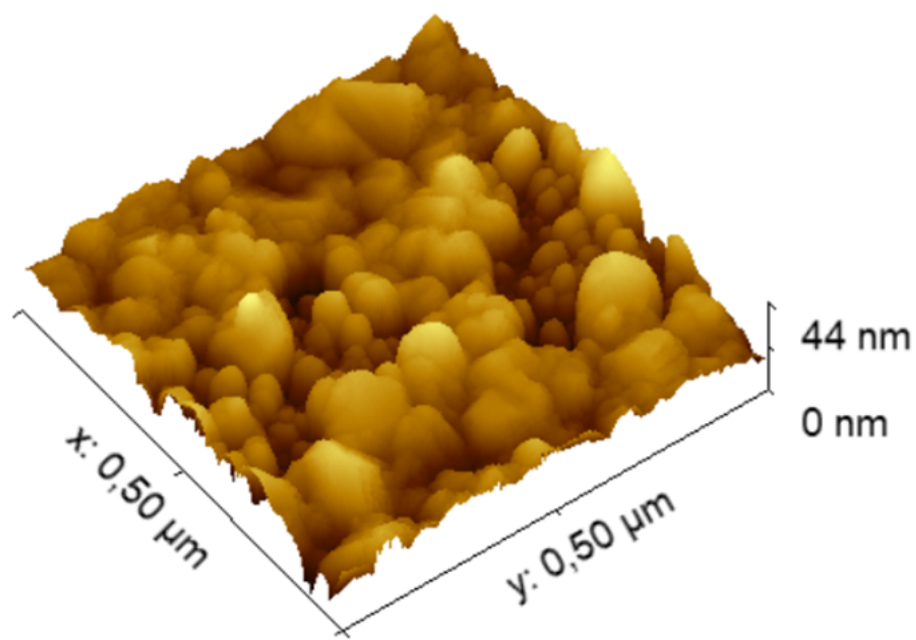

*Figure S5* 3D-image of the TD silver surface obtained by AFM. The image shows the irregular nanostructure of the Ag layer

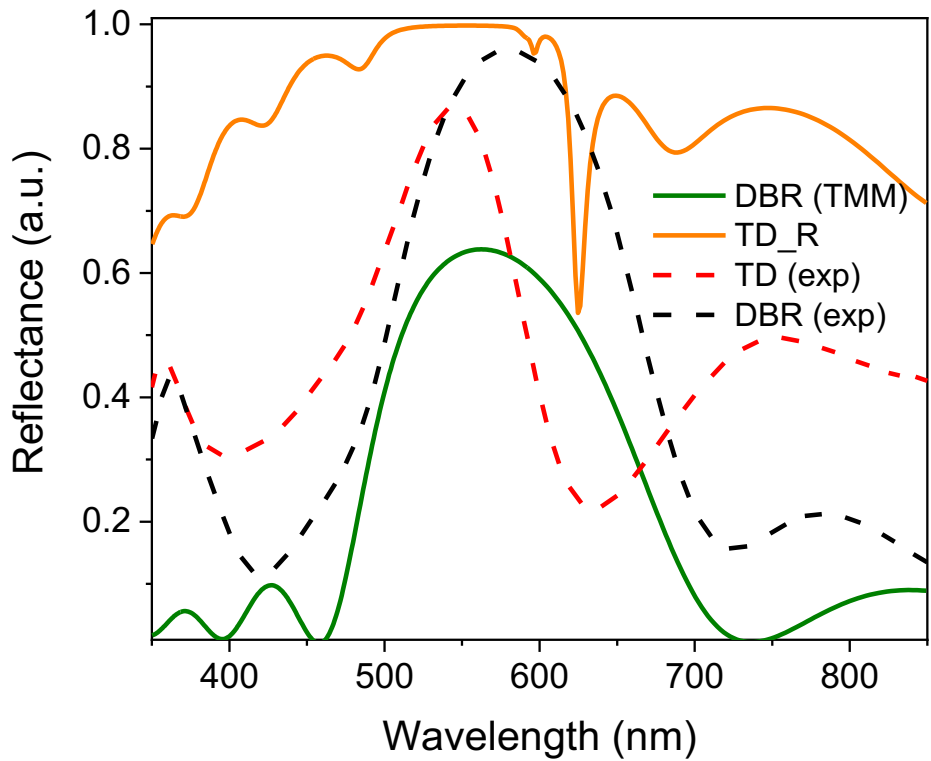

*Figure S6* Reflectance spectra showing the experimental data of the TD (red dotted line) and DBR (black dotted line), together with the simulated curve of the TD (orange line) and DBR (green line) calculated with the Transfer Matrix Method (TMM) reported in the Matlab code section of the supplementary materials. The spectral positions of both Tamm feature and photonic bandgap align with experimental result while the shape of the TD simulation resembles the case of bulk structure Ag layer.

**Optical simulation of the TP device**

Transfer Matrix Method (TMM), implemented in MATLAB, has been executed modeling the optical response of the DBR and the TD to optimize and validate the experimental data. Due to their nanoporous nature of the layers also observed by AFM analysis, the TD and the DBR have been modeled applying the Bruggeman effective medium approximation. In the simulation (Fig. S6), the best agreement with experimental data was obtained using a $SiO_2$ layer thickness of 110 nm and a $TiO_2$ layer thickness of 90 nm, with a filling factor of 0.57 applied to both layers to account for porosity. For the TD, the optimal model included a Ag layer thickness of 33 nm and a filling factor of 0.82. This approach yielded a spectral alignment between the simulated and experimental DBR and TD features, although the simulated Tamm resonance appeared spectrally narrower than its experimental counterpart.

## Matlab Code

```
clc; clear; close all;
%% Starting Parameters

% experimental data to compare with the model
% load('DBR_R.mat');
% wlPC=DBR_R(:,1);
% wlPC=table2array(wlPC);
% RPC=DBR_R(:,2);
% RPC=table2array(RPC)./100;
% load('TD_R.mat');
% wlAgPC=TD_R(:,1);
% wlAgPC=table2array(wlAgPC);
% RAgPC=TD_R(:,2);
% RAgPC=table2array(RAgPC)./100;

%plasmonic nanocrystals Ag Carsten Soennichsen PhD thesis Land./B¨orn.: collection of values by Foiles (1985).

eps_m=[1]; %air - j parameter
NC1=5.76*10^28; %carrier density m^-3
m_h1=0.96*(9.1e-31); %
eps_i=3.7;
e=1.6e-19;%C
eps0=8.85e-12;
omega_p1=((NC1*(e*e))/(eps0*m_h1)).^0.5;%1/s
E_p1=6.5821e-16*omega_p1;%eV
```

```
gamma=0.018; %eV -

n0=1;

ns=1.45;

% Material and geometric parameters

fv=0.82; %filling factor Ag

fv2=0.57; %filling factor TiO2

fv3=0.57; %filling factor SiO2

d1=33; %Ag thickness [nm]

d2=90; %titanium dioxide layer thickness [nm]

d3=110; %silicon dioxide layer thickness [nm]

%% TMM

for j=1:1

ii=0;

for i=1:0.005:4.5

ii=ii+1;

l(ii)=i;

%plasmonics Ag

eps1_1_Ag(j,ii)=eps_i-((E_p1.^2)/((l(ii)^2)+(gamma^2)));

eps2_1_Ag(j,ii)=(((E_p1.^2)*gamma)/(l(ii)*((l(ii)^2)+(gamma^2))));

eps1_Ag(j,ii)=complex(eps1_1_Ag(j,ii),eps2_1_Ag(j,ii));

%Effective medium approx. for Ag film dielectric function

eps_mean1(j,ii) = eps_m(j)*(2*(1-fv)*eps_m(j)+(1+2*fv)*eps1_Ag(j,ii))/((2+fv)*eps_m(j)+(1-fv)*eps1_Ag(j,ii));

n_Ag(j,ii)=conj(sqrt(eps_mean1(j,ii)));

xm(ii)=1240/l(ii)*1e-7;

% Titanium Oxide in micrometers extrapolated from F. Scotognella et al. / Ceramics International 41 (2015) 8655–8659

ll(ii)=1240./(l(ii)*1000); %[um]

n2(ii)=sqrt(4.99+...
```

```
1/(96.6*(ll(ii)^1.1))+...
1/(4.6*(ll(ii)^1.95)));
eps2(ii)=n2(ii).^2;
yf(ii)=eps_m(j)*(2*(1-fv2)*eps_m(j)+(1+2*fv2)*eps2(ii))/((2+fv2)*eps_m(j)+(1-fv2)*eps2(ii));
neT(ii)=sqrt(yf(ii));
%n SiO2 extrapolated from refractiveindex.info (Malitson 1965)
n3(ii)=sqrt(1+...
(0.6961663*ll(ii)^2/(ll(ii)^2-0.0684043^2))+...
(0.4079426*ll(ii)^2/(ll(ii)^2-0.1162414^2))+...
(0.8974794*ll(ii)^2/(ll(ii)^2-9.896161^2)));
eps3(ii)=(n3(ii)).^2;
yff(ii)=eps_m(j)*(2*(1-fv3)*eps_m(j)+(1+2*fv3)*eps3(ii))/((2+fv3)*eps_m(j)+(1-fv3)*eps3(ii));
neS(ii)=sqrt(yff(ii));
%Ag layer
S1(j,ii)=cos(2*3.14*(n_Ag(j,ii))*d1*l(ii)/1240);
P1(j,ii)=(1i*sin(2*3.14*(n_Ag(j,ii))*d1*l(ii)/1240))/(n_Ag(j,ii));
Q1(j,ii)=1i*(n_Ag(j,ii))*sin(2*3.14*(n_Ag(j,ii))*d1*l(ii)/1240);
R1(j,ii)=cos(2*3.14*(n_Ag(j,ii))*d1*l(ii)/1240);
M1=[S1(j,ii), P1(j,ii); Q1(j,ii), R1(j,ii)];
%Titanium dioxide
S2(j,ii)=cos(2*3.14*neT(ii)*d2*l(ii)/1240);
P2(j,ii)=(1i*sin(2*3.14*neT(ii)*d2*l(ii)/1240))/neT(ii);
Q2(j,ii)=1i*neT(ii)*sin(2*3.14*neT(ii)*d2*l(ii)/1240);
R2(j,ii)=cos(2*3.14*neT(ii)*d2*l(ii)/1240);
M2=[S2(j,ii), P2(j,ii); Q2(j,ii), R2(j,ii)]; %matrix of TiO2 monolayer
%Silicon dioxide
S3(j,ii)=cos(2*3.14*neS(ii)*d3*l(ii)/1240);
P3(j,ii)=(1i*sin(2*3.14*neS(ii)*d3*l(ii)/1240))/neS(ii);
Q3(j,ii)=1i*neS(ii)*sin(2*3.14*neS(ii)*d3*l(ii)/1240);
R3(j,ii)=cos(2*3.14*neS(ii)*d3*l(ii)/1240);
M3=[S3(j,ii), P3(j,ii); Q3(j,ii), R3(j,ii)]; %matrix of SiO2 monolayer
%product Distributed Bragg reflactors
```

```matlab
MM=((M2*M3)^5);
t1=2*ns/(((MM(1,1)+MM(1,2)*n0)*ns)+(MM(2,1)+MM(2,2)*n0));
T1(j,ii)=(n0/ns)*(abs(t1)^2);
r=(((MM(1,1)+MM(1,2)*n0)*ns)-(MM(2,1)+MM(2,2)*n0))/...
(((MM(1,1)+MM(1,2)*n0)*ns)+(MM(2,1)+MM(2,2)*n0));
Re1(ii)=(abs(r)^2);
%product TD
MM2=((M2*M3)^5)*M1;
t2=2*ns/(((MM2(1,1)+MM2(1,2)*n0)*ns)+(MM2(2,1)+MM2(2,2)*n0));
T2(j,ii)=(n0/ns)*(abs(t2)^2);
r2=(((MM2(1,1)+MM2(1,2)*n0)*ns)-(MM2(2,1)+MM2(2,2)*n0))/...
(((MM2(1,1)+MM2(1,2)*n0)*ns)+(MM2(2,1)+MM2(2,2)*n0));
Re2(ii)=(abs(r2)^2);
Abs1(j,ii)=-log10(T1(j,ii));
lambda(ii)=1240/l(ii); %[nm]
end
end

%% Grapichs

%Graph combining experimental and simulated data
% h1=figure;
% h2=plot(lambda,Re1,lambda,Re2,wlPC,RPC,wlTD,RATD);
% set(h2(1),'Color',[0 0 0],'LineWidth',2);
% set(h2(2),'Color',[0 0 1],'LineWidth',2);
% set(h2(3),'Color',[1 0 0],'linestyle','-.','LineWidth',2);
% set(h2(4),'Color',[0 1 0],'linestyle','-.','LineWidth',2);
% %set(h1, 'Position', [1030 400 550 480]);
% set(h2,'linewidth',2);
% set(gca,'fontsize',14,'FontWeight','bold');
% title([num2str(d1) ' nm of Silver, sample 5R']);
% xlabel('Energy (eV)','FontSize',18,'FontWeight','bold');%x-axis label
```

```
% ylabel('R','FontSize',18,'FontWeight','bold');%y-axis label
% axis([320 760 -0.05 1.02]);
% legend('TMM DBR','TMM TD','Exp DBR','Exp TD');

% graph of the model
h1=figure;
h2=plot(lambda,Re1,lambda,Re2);
set(h2(1),'Color',[0 0 0],'LineWidth',2);
set(h2(2),'Color',[0 0 1],'LineWidth',2);
%set(h1, 'Position', [1030 400 550 480]);
set(h2,'linewidth',2);
set(gca,'fontsize',14,'FontWeight','bold');
title([num2str(d1) ' nm of Silver on DBR']);
xlabel('Wavelength (nm)','FontSize',18,'FontWeight','bold');%x-axis label
ylabel('R','FontSize',18,'FontWeight','bold');%y-axis label
axis([320 760 -0.05 1.02]);
legend('TMM of DBR','TMM of TD');
```